\documentclass[11pt]{article}
\usepackage{epsfig}
\usepackage{graphicx}
\usepackage{amsmath}
\usepackage{amssymb}

\def\stand{{\rm stand}}

\begin{document}
\vspace{32pt}

\begin{center}

\centerline{\textbf{\Large
Atmospheric lepton fluxes at very high energy}}

\vspace{40pt}

J.I.~Illana$^1$, P.~Lipari$^2$, M.~Masip$^1$,
D.~Meloni$^3$
\vspace{15pt}

{\it $^1$CAFPE and
Depto.~de F{\'\i}sica Te\'orica y del Cosmos, Universidad de Granada,
18071 Granada, Spain}

\vspace{10pt}

{\it $^2$INFN, Universit\`a degli Studi di Roma ``La Sapienza",
00185 Rome, Italy}

\vspace{10pt}

{\it $^3$Institut f\"ur Theoretische Physik und Astrophysik,
Universit\"at W\"urzburg, 97074 W\"urzburg, Germany}

\vspace{15pt}

\texttt{jillana@ugr.es, paolo.lipari@roma1.infn.it, masip@ugr.es,
davide.meloni@physik.uni-wuerzburg.de}

\end{center}

\vspace{10pt}

\begin{abstract}
The observation of astrophysical neutrinos requires a detailed
understanding of the atmospheric neutrino background. Since
neutrinos are produced in meson decays together with a charged
lepton, important constraints on this background can be obtained
from the measurement of the atmospheric muon flux. Muons, however,
can also be produced as $\mu^+ \mu^-$ pairs by purely
electromagnetic processes. We use the $Z$--moment method to study
and compare the contributions to the atmospheric muon and neutrino
fluxes from different sources ($\pi$/$K$ decay, charmed and
unflavored hadron decay, and photon conversion into a muon pair). 
We pay special attention to the contribution from unflavored mesons
($\eta$, $\eta^\prime$, $\rho^\circ$, $\omega$ and $\phi$). 
These mesons are abundant in air showers, 
their lifetimes are much shorter than those of 
charged pions or kaons, and they have decay branching ratios of 
order $10^{-4}$ into final states containing a muon pair.
We show that they may be the
dominant source of muons at $E_\mu \gtrsim 10^3$ TeV. 
\end{abstract}

\newpage

\section{Introduction}
Atmospheric muons and neutrinos are produced in the showers of
high energy cosmic rays in the Earth's atmosphere. Neutrinos can
change flavor during their travel from the creation to the
detection points, and therefore the observation of their fluxes
allows to study their masses, mixings and interactions. Above a
minimum energy of a few GeV muons reach the ground, where they
can also be observed. Measurements of the atmospheric muon flux
provide information about the primary cosmic ray fluxes and about
the properties of high-energy hadronic interactions. Moreover,
the measurements of the atmospheric muons can be used to obtain
precise estimates of the fluxes of $\nu_\mu$ and
$\overline{\nu}_\mu$ {\em before} flavor oscillation effects. In
fact, these measurements have played an important role for the
interpretation of the data 
and the determination of the flavor oscillation parameters. 
The dominant source of muons and muon neutrinos is
the weak decay of charged pions and kaons. These decays always
produce $\ell^+ \nu_\ell$ or $\ell^- \overline{\nu}_\ell$ pairs,
implying a correlation between charged-lepton and neutrino fluxes
that can be robustly predicted.

An important goal of present and future experiments is the
detection of high-energy neutrinos produced in galactic and
extragalactic astrophysical sources \cite{Gaisser:1994yf,Lipari:2006uw}.
Neutrinos propagate without
significant losses (excluding the cosmological redshift) from
very distant sources, and one of largest expected signals 
is an isotropic diffuse flux generated by the ensemble of all
extragalactic sources in the universe. For neutrino astronomy,
therefore, the atmospheric neutrino fluxes constitute a disturbing
background that must be subtracted from the signal. Measurements
of the muon flux can help in the determination of this background.

At very high energy it is possible, and indeed virtually certain,
that the decay of charged pions and kaons does not remain the
dominant source of atmospheric muons and neutrinos. The reason
is that these particles are relatively long lived and, because
of the Lorentz time dilatation, their decay probability at high
energy is strongly suppressed. As an illustration, the decay
length of a 10~TeV charged pion is more that 500~Km, around 100
times larger than its interaction length in air. The contribution
of particles with a shorter lifetime is therefore likely to
become dominant. In particular, the contribution of charmed
hadrons is a natural candidate. These particles have large ($\sim
10$\%) branching ratios into semileptonic modes and a
lifetime $\tau \sim 10^{-12}$~s, implying a decay probability of
order 1 up to energies around $10^{7}$~GeV.

Recently, however, it has been suggested \cite{Illana:2009qv}
that the dominant source of atmospheric muons of very large
energy could be the electromagnetic decay of unflavored mesons
into $\mu^+\mu^-$ pairs. These mesons are particles of type
$q_f \overline{q}_f$, constituted by a quark and an anti--quark
of the same flavor. Neglecting heavy quarks one has 3 scalar
($\pi^\circ$, $\eta$ and $\eta^\prime$) and three vector
($\rho^\circ$, $\omega$ and $\phi$) mesons of this kind. They decay
mostly into pions and photons via strong or electromagnetic
interactions with a very short lifetime. All of them except the
neutral pion (which is below threshold) can decay into a
$\mu^+\mu^-$ pair, sometimes together with a photon or a neutral
pion ($\eta \to \mu^+\mu^-\gamma$ or $\omega \to
\mu^+\mu^-\pi^\circ$), with small branching ratios of order
$10^{-4}$. These rare decay modes have been neglected in
essentially all calculations of atmospheric muons. However, 
if the multiplicities and the energy spectra of
charged pions and unflavored mesons are roughly similar, then
they will become the dominant 
source of atmospheric muons when the average decay
probability of charged pions is suppressed by a factor ($\sim
10^{-4}$) of the same order as 
the branching ratios into the rare modes
that contain muons.

In this article we do a critical review of the different
contributions to the lepton fluxes at high energies. 
In Section 3 we estimate the 
conventional lepton fluxes using the so called 
$Z$--moment method, that provides simple analytic
expressions.
Then we focus on the contribution to the muon flux from the
decay of unflavored mesons. In
sections 5 and 6 we evaluate, respectively, the contribution from
the prompt decay of charmed hadrons and from $\gamma$ conversion
into a $\mu^+\mu^-$ pair. Finally, 
we summarize the uncertainties and the
implications of our results.

\section{Components of the atmospheric lepton fluxes}
The atmospheric flux of a lepton type $\ell$ can be described
as the sum of contributions from the decay of different unstable
particles:
\begin{equation}
\phi_\ell (E, \theta) = \sum_j \phi_\ell^{(j)} (E, \theta)\,.
\label{eq:ell}
\end{equation}
In this equation $E$ is the energy and $\theta$ the zenith angle
of the lepton, and the summation runs over all possible parent
particles. 
In this work with {\it lepton} we refer 
only to neutrinos and muons. The production of electrons and positrons is
dominated by photon conversion in the electromagnetic field of
the air nuclei ($\gamma Z \to e^+ e^- Z$, with $Z$ the electric charge
of the nucleus) and will not be
discussed here.

The parent particles that are the source of atmospheric leptons
can be naturally divided in three classes. The first class
(standard contribution)
includes charged pions and kaons that decay via charged-current
weak interactions into lepton pairs: ($e^+\nu_e$), 
$(\mu^+\nu_\mu)$ and the charge conjugate states. The observed 
atmospheric fluxes can at the present time be
entirely attributed to this standard contribution.

The second source of atmospheric leptons (charm contribution)
is the weak decay of particles 
that contain a charm (anti)--quark. These decays 
also generate leptons in $(e\nu_e)$ and $(\mu\nu_\mu)$
pairs. In addition, $D_s^\pm$ mesons (scalar mesons with a
$c\overline{s}$ or $s\overline{c}$ quark content) 
have a decay branching ratio of $\sim 6.4$\% into 
 the 2--body mode
$\tau^+\nu_\tau$ ($\tau^-\overline{\nu}_\tau$);
the subsequent decay of the $\tau$ lepton generates a second
tau (anti)--neutrino.
This chain decay process is the main source of atmospheric 
$\nu_\tau$ and $\overline{\nu}_\tau$.
The contribution of charmed particles to
the atmospheric lepton fluxes is subdominant and currently
undetected. 
It is however expected that this mechanism 
will overtake the standard contribution
at sufficiently high energy.

A third class of parent particles 
can contribute to the flux 
of atmospheric muons (but not of neutrinos).
This {\it unflavored contribution} is due to the
decay of the unflavored mesons $\eta$,
$\eta^\prime$, $\rho^\circ$, $\omega$ and $\phi$. These particles
have small (order $10^{-4}$) branching ratios into final states that include a
$\mu^+\mu^-$ pair. The possible significance of this
contribution has been discussed in \cite{Illana:2009qv}, and it
will be critically analyzed in the following.

Muon pairs can also be directly produced in Drell--Yan
processes and in photon conversions of type 
$\gamma Z \to \mu^+ \mu^- Z$.
This last process,
despite being suppressed by a factor $(m_e/m_\mu)^2 \sim 2.3\times 10^{-5}$
with respect to the production of $e^+e^-$ pairs,
is potentially interesting.
Its contribution to the muon atmospheric flux
will be indicated as $\phi_\mu^{(\gamma)}$ and 
discussed later.

Taking into account these different sources, 
the muon and neutrino fluxes can then be expressed as the sum of 
four and two components, respectively:
\begin{eqnarray}
\phi_{\nu_\alpha} (E, \theta) & = &
\phi_{\nu_\alpha}^\stand (E, \theta)
+
\phi_{\nu_\alpha}^{\rm charm} (E, \theta)\;.
\label{eq:lep1}
\\[0.22 cm]
\phi_{\mu} (E, \theta)& = &
\phi_{\mu}^\stand (E, \theta)
+
\phi_{\mu}^{\rm charm} (E, \theta)
+
\phi_{\mu}^{\rm unflav} (E, \theta)
+
\phi_{\mu}^{(\gamma)} (E, \theta) ~.
\label{eq:lep2}
\end{eqnarray}

\section{The standard contribution} 
Simple analytic expressions for the atmospheric 
lepton fluxes produced by the decay
of charged pions and kaons are described in the 
textbook \cite{gaisser} by Gaisser 
(see \cite{lipari93} for additional details).
These expressions are obtained under 
three simplifying assumptions:
\begin{itemize}
\item[(i)] The interaction lengths $\lambda_k$ of
all hadrons (labeled by $k$) are taken constant, 
neglecting their energy dependence.
\item[(ii)] The inclusive spectra of
secondary particles $j$ created by 
the projectile particle $k$ in a hadronic interaction 
with an air nucleus satisfy the scaling condition:
\begin{equation}
\frac{dn_{kj}}{dE} (E_j; E_k) \equiv \frac{1}{\sigma_k} \;
\frac{d\sigma_{kj}}{dE} (E_j; E_k) \simeq \frac{1}{E_k} \;
F_{kj} (x)\,, \label{eq:scal0}
\end{equation}
where $x=E_j/E_k$, $\sigma_k$ is the total 
inelastic cross section 
and $F_{kj}(x)$ is the number density of particles 
$j$ carrying a fraction $x$ of the initial energy
after the collision. 
\item[(iii)] The primary nucleon fluxes 
are simple power laws of exponent $\alpha$:
\begin{equation}
\begin{aligned}
\phi_p (E_0) = & K ~p_0 ~E_0^{-\alpha}\,;
\\[0.2 cm]
\phi_n (E_0) = & K ~n_0 ~E_0^{-\alpha} = K ~(1-p_0)
~E_0^{-\alpha}\,.
\end{aligned}
\end{equation}
\end{itemize}

In the {\it low-energy limit}, when the parent particle 
(a charged pion or kaon) decays with 
probability close to one, the lepton fluxes are isotropic 
and have a power-law energy spectrum 
with the same exponent $\alpha$ as the primary nucleon fluxes.
The component $\phi_{\ell}^{(j)}$ takes then the form 
\begin{equation}
\phi_{\ell}^{(j)} (E, \theta) = K ~E^{-\alpha} \, 
A_{j} (\alpha) ~Z_{j \ell} (\alpha) \, .
\label{eq:form1}
\end{equation}
$A_j(\alpha)$ is 
the ratio between the number of nucleons that reach the 
Earth with energy in the
interval ($E$,$E+dE$) and the number of 
particles of type $j$ produced in the same energy 
interval by primary or secondary particles.
This quantity is less than one even 
if primary nucleons tend to generate
many secondary particles of type $j$, because these 
particles are 
produced with lower energy while the ratio is 
performed at a fixed $E$.
The second quantity in Eq.~(\ref{eq:form1}), 
$Z_{j\ell} (\alpha)$, is analogous, it relates
the lepton flux with the flux of its parent particle $j$.
It can be calculated as the 
($\alpha-1)$--moment of the inclusive spectrum 
$F_{j\ell} (x)$ of
lepton $\ell$ from the decay of $j$:
\begin{equation}
Z_{j\ell} (\alpha) = \int_0^1 dx ~x^{\alpha -1} ~ F_{j\ell} (x) \, ,
\label{eq:Z-def}
\end{equation}
where $x = E_\ell/E_j$ and $F_{j\ell} (x)$ is 
taken in any frame where the
parent particle is ultrarelativistic.
The quantity $A_j(\alpha)$ includes 
proton and neutron contributions:
\begin{equation}
A_j (\alpha) =
 p_0 \, A_{pj} (\alpha) + n_0 \, A_{nj} (\alpha)\;. 
\end{equation}
It is straightforward (see \cite{gaisser,lipari93})
to obtain these contributions in terms of $Z$--factors:
\begin{equation}
A_{pj} \pm A_{nj} = \frac{Z_{pj} \pm Z_{nj}} {1 - Z_{pp} \mp Z_{pn}} \,,
\label{eq:defA}
\end{equation}
where the dependence on $\alpha$ is
implicit. Again, the $Z$--factor
$Z_{kj}(\alpha)$ is just the $(\alpha-1)$--moment 
of $F_{kj} (x)$, 
\begin{equation}
Z_{kj} (\alpha) = \int_0^1 dx ~x^{\alpha -1} ~ F_{kj} (x)\,.
\label{eq:Z-def1}
\end{equation}

\begin{figure} [ht]
\begin{center}
\includegraphics[width=16.0cm]{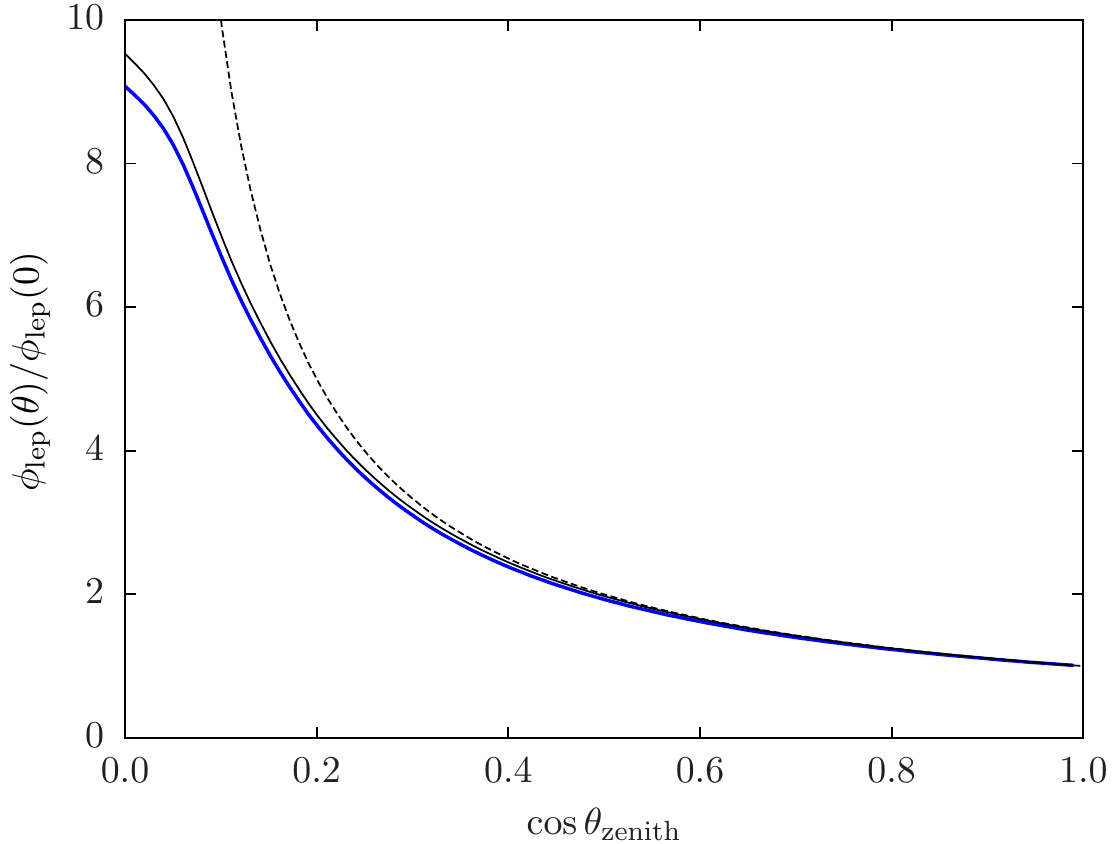}
\end{center}
\caption {\footnotesize
Zenith angle distribution of 
very high energy lepton fluxes
from pion  and kaon  decay (thick solid line).
The dashed line is $(\cos\theta)^{-1}$.
The thin solid line represents the   curve $[\cos \theta^*(\theta)]^{-1}$,
where $\theta^* (\theta)$ is the local zenith angle (the angle with the
vertical direction)  at the point  that corresponds to the  column  density
$t = 200$~(g~cm)$^{2}$  for  the line  of sight  defined by 
zenith angle  $\theta$  at sea level.
\label{fig:zenith} }
\end{figure}

At high energy the decay probability of pions and kaons 
is suppressed because of the 
Lorentz time dilatation.
When the decay probability of the parent particle is small,
using the 
assumptions (i), (ii) and (iii) it is possible to express 
the lepton flux from $j$--decay as
\begin{equation}
\phi_{\ell}^{(j)} (E, \theta)
 = 
 K ~E^{-\alpha}
~ \frac{\varepsilon_j \, F_{\rm zenith} (\theta)} {E} ~ B_{j} (\alpha)
~Z_{j\ell} (\alpha+1)\,.
\label{eq:form2}
\end{equation}
This energy spectrum is also a power law, 
but its slope is a unit steeper than in the primary nucleon flux.
In addition, it has the strong dependence 
on the zenith angle 
described by $F_{\rm zenith}(\theta)$
(shown in Fig.~\ref{fig:zenith}).
For $\theta \lesssim 60^\circ$ this function is well 
approximated by a ``secant law'', 
\begin{equation}
F_{\rm zenith} (\theta) \simeq \frac{1}{\cos\theta} \;,
\end{equation}
whereas for larger zenith angles it keeps growing
monotonically, reaching at $\theta \simeq 90^\circ$ a 
value close to 10.
To a good approximation the zenith angle dependence 
obtained in this {\it high energy limit} is
universal, it does not depend 
on the parent particle type 
or the details of the hadronic interactions.
The quantity $\varepsilon_j$ is the {\it critical energy} 
for particle $j$:
\begin{equation}
\varepsilon_j = \frac{ h_0 \, m_j}{ c\, \tau_j} 
\end{equation}
that corresponds to the condition  where the  decay length is equal to 
the scale height of the air density in the stratosphere  
$h_0 \simeq 6.36$~Km ($\rho (h) \propto e^{-h/h_0}$).
The critical energies for $\pi^\pm$, $K_L$ and $K^\pm$
are approximately 115, 210 and 850 GeV, respectively.
After averaging over the creation position,
the decay probability of  a particle of type  $j$ 
energy $E$  and zenith angle $\theta$, for  large energy takes 
the asymptotic form:
\begin{equation}
P_{\rm dec}  =   \frac{\varepsilon_j \; 
F_{\rm zenith} (\theta)} {E} ~\frac{B_j(\alpha)}{A_j(\alpha)}
\simeq   \frac{\varepsilon_j \;  F_{\rm zenith} (\theta)} {E} 
\label{eq:pdec}
\end{equation}
Note also that 
the decay $Z$--factor in Eq.~(\ref{eq:form2}) is 
calculated for the argument $(\alpha+1)$.
Finally, $B_j(\alpha)$ is analogous to the quantity 
$A_j(\alpha)$ defined in the low-energy limit
but includes effects due to the distribution 
of the creation point of particle $j$.
It can be separated as
\begin{equation}
B_j (\alpha) =
 p_0 \, B_{pj} (\alpha) + n_0 \, B_{nj} (\alpha)\;.
\end{equation}
Including only the particles produced in nucleon
interactions one obtains:
\begin{equation}
B_{pj} (\alpha) \pm B_{nj} (\alpha) = 
\frac{Z_{pj} \pm Z_{nj}}{1-Z_{pp} \mp Z_{pn}} 
\; \left ( \frac{\lambda_j}{\lambda_j - \Lambda_N^\pm} \right ) 
~\ln \left ( \frac{\lambda_j}{\Lambda_N^\pm} \right )\;,
\end{equation}
where the dependence on $\alpha$ is implicit and 
\begin{equation}
\Lambda_N^\pm = \frac{\lambda_N}{1 -Z_{pp} \mp Z_{pn}}\;.
\end{equation}
For pions, the inclusion of {\em regeneration} effects 
(the contribution of pions produced in pion interactions)
yields the result:
\begin{equation}
B_{p\pi^\pm} (\alpha) \pm B_{n\pi^\pm} (\alpha) = 
\frac{Z_{pj} \pm Z_{nj}}{1-Z_{pp} \mp Z_{pn}} 
\; \left ( \frac{\Lambda_\pi^\pm}{\Lambda_\pi^\pm
- \Lambda_N^\pm} \right ) ~\log \left (
\frac{\Lambda_\pi^\pm}{\Lambda_N^\pm} \right )\;,
\end{equation}
with:
\begin{equation}
\Lambda_\pi^\pm = \frac{\lambda_\pi}{1 -Z_{\pi^+\pi^+} \mp Z_{\pi^+\pi^-}}\;.
\end{equation}
Analogous expressions for kaons that 
include the effect of regeneration are
discussed in \cite{lipari93}.

Eq.~(\ref{eq:form2}) has been obtained 
under the hypothesis that the decay probability
of the parent particle $j$ is small, {\it i.e.},
$E \gg \varepsilon_j \; F_{\rm zenith} (\theta)$.
Correspondingly, the range of validity of the flux in
Eq.~(\ref{eq:form1}) is
$E \ll \varepsilon_j \; F_{\rm zenith} (\theta)$. 
A useful expression that interpolates between 
the asymptotic fluxes in (\ref{eq:form1}) and (\ref{eq:form2}) 
is:
\begin{equation}
\phi_{\ell}^{(j)} (E, \theta) = \left ( K ~E^{-\alpha}
\right ) ~ A_j (\alpha) ~Z_{j \ell} (\alpha) ~\left [ 1 +\frac{E} 
{\varepsilon_j \; F_{\rm zenith} (\theta)} 
\frac{A_j (\alpha)}{B_j(\alpha)}
~\frac{Z_{j\ell} (\alpha)}{Z_{j \ell} (\alpha+1)} \right ]^{-1}\,.
\label{eq:form3}
\end{equation}

The analytic expressions described above have 
a limited validity, 
since assumptions (i), (ii) and (iii) 
are not rigorously correct.
The hadronic cross sections grow slowly with energy;
the energy scaling (\ref{eq:scal0}) is not exact; and 
the high-energy cosmic ray flux cannot be described as
a simple power law because of the steepening at the 
cosmic ray {\it knee} 
($E_{\rm knee} \simeq 3 \times 10^{6}$~GeV).
In a first approximation,
the energy dependence of hadronic interaction
lengths and the violation of scaling
in inclusive particle distributions can be taken into account
just by considering the 
quantities $\lambda_k$ and $Z_{kj} (\alpha)$ as (slowly 
varying) functions of the 
lepton energy. The calculation of the lepton fluxes 
for an arbitrary shape
of the primary flux is discussed in Subsection~\ref{sec:yields}.

\subsection{Numerical estimate}
The objective in 
this work is to compare the different contributions to the lepton 
fluxes at very high energy
($E_\ell \gtrsim 10$~TeV).
To estimate the standard contribution from pion and 
kaon decays one can therefore
use Eq.~(\ref{eq:form2}). 
Summing over all parent particles
this contribution (for zenith angles $\theta \lesssim 60^\circ$)
takes the form:
\begin{equation}
\frac{\phi_\ell^\stand (E, \theta)}{\left ( K \; E^{-\alpha}
\right )} \simeq \frac{ \mathbf{E}_\ell (\alpha)}{E \; \cos
\theta}\,,
\label{eq:echar_def}
\end{equation}
where the constant $\mathbf{E}_\ell (\alpha)$ has dimension of
energy and is given by
\begin{equation}
\mathbf{E}_\ell (\alpha) = \sum_{j \in \{\pi^\pm, K^\pm,K_L \}}
\varepsilon_j ~B_j(\alpha) ~Z_{j \ell } (\alpha +1)\,.
\end{equation}

For a numerical estimate of the 
quantities $\mathbf{E}_\ell (\alpha)$, we first have obtained 
the inclusive particle spectra from 
a MonteCarlo simulation generated by the code Sibyll \cite{sibyll}, 
and then we have 
deduced the hadronic $Z$--factors from numerical integrations. 
The interaction lengths in air have been taken
from the PDG fit \cite{PDG}
of hadron--nucleon cross sections, and we have used a 
Glauber formalism \cite{glauber} to 
compute the cross sectios on a nuclear target. 
The hadronic $Z$--factors at 
$E_\ell \simeq 10^{6}$~GeV that we obtain are 
shown in Table~\ref{tab:echar}.
\begin{table} [bt]
\begin{center}
\begin{tabular}{| c | c | c | }
\hline $\mathbf{E}_\ell\; [{\rm GeV}]$
& $\alpha = 2.7$ & $\alpha = 3.0$ \\
\hline
$(\mu^+ + \mu^-) $ & 9.3 & 4.9 \\
$(\nu_\mu + \overline{\nu}_\mu ) $ & 3.4 & 1.7 \\
$( \nu_e + \overline{\nu}_e ) $ & 0.17 & 0.10 \\
$\mu^+ $ & 5.2 & 2.8 \\
$\mu^- $ & 4.1 & 2.1 \\
$\nu_\mu $ & 2.2 & 1.1 \\
$\overline{\nu}_\mu $ & 1.2 & 0.6 \\
$\nu_e $ & 0.10 & 0.06 \\
$\overline{\nu}_e $ & 0.07 & 0.04 \\
\hline
\end{tabular}
\end{center}
\caption{Characteristic energies 
$\mathbf{E}_\ell$ (see definition in Eq.~(\protect\ref{eq:echar_def}))
for the lepton fluxes
calculated using the
Sibyll interaction model.
The quantities are evaluated for $E_\ell \simeq 10^{6}$~GeV.
\label{tab:echar}
}
\end{table}
The characteristic energy for muons, summing over both charges, is
\begin{equation}
\begin{aligned}
& \mathbf{E}_\mu(\alpha = 2.7) \simeq 9.3 ~{\rm GeV} \; ,
\\[0.21 cm]
& \mathbf{E}_\mu(\alpha = 3.0) \simeq 4.9 ~{\rm GeV}\; .
\end{aligned}
\label{eq:num_standard}
\end{equation}
These estimates are given for the value of $\alpha$ 
relevant below and above the cosmic-ray knee.

The energy dependence of the inclusive particle spectra in 
the fragmentation
region (that dominates the integrand of the $Z$--factor 
expression) in the Sibyll code is weak,
and the characteristic energies for $E_\ell \simeq 10^{6}$~GeV 
(given in Table~\ref{tab:echar})  
are approximately 25\% lower than the results obtained at
$E_\ell \simeq 10^{4}$~GeV, where the calculation 
is in reasonable agreement with observations.
Because of our poor understanding of hadronic interactions
one should attribute around $\sim 30$\% of 
systematic uncertainty to the numerical estimates in 
Table~\ref{tab:echar} and
Eq.~(\ref{eq:num_standard}), but a larger error 
cannot be excluded.
 
Finally we note that the absolute prediction for the
lepton fluxes   also depends on the estimate of the primary nucleon 
flux \cite{tibet_spectrum:2008jb,Abbasi:2007sv,auger_spectrum_2010}.
This introduces perhaps the largest uncertainty, 
because the cosmic-ray composition 
above the knee is poorly known.

\subsection{Lepton yields}
\label{sec:yields}
To compute the lepton fluxes when the primary spectrum 
 is {\it not} a power law
it is necessary to introduce the lepton yields.
The lepton yield $Y_{p(n) \to \ell}(E, E_0, \theta)$
of a primary nucleon of energy 
$E_0$ and zenith angle $\theta$
gives the average number of leptons $\ell$ of
energy $E$ observable at ground level
(where to a good approximation the shower has completed its 
development) per unit energy:
\begin{equation}
Y_{p(n) \to \ell} (E, E_0, \theta) \equiv \frac{dN_{p(n) \to
\ell}}{dE} (E, E_0, \theta)\,.
\end{equation}
The lepton flux can then be calculated with a simple integration:
\begin{equation}
\phi_\ell (E, \theta) = \int_E^\infty dE_0 ~ \left [ \phi_p (E_0) \;
Y_{p \to \ell} (E, E_0, \theta) + \phi_n (E_0) \; Y_{n \to \ell} (E,
E_0, \theta) \right ]\,. 
\label{eq:inty}
\end{equation}
$Y_{p(n) \to \ell}$ can be written as a sum of 
terms associated to the production 
and decay of different parent particles:
\begin{equation}
Y_{p(n) \to \ell} (E, E_0, \theta) = \sum_j Y_{p(n) \to
\ell}^{(j)} (E, E_0, \theta)\,,
\end{equation}
where the index $j$ runs over all species with decay modes
containing the lepton $\ell$.

Using assumptions (i) and (ii) these yields take 
simple scaling forms. In the low-energy limit 
described in the previous section one has
\begin{eqnarray}
Y_{N \to \ell}^{(j)} (E, E_0, \theta) & = & \frac{1}{E_0} ~
Q_{N\mu}^{(j)} \left( x \right)\nonumber
 = \frac{1}{E_0} ~ \left[ G_{Nj} \otimes F_{j\ell}
\right]\left( x \right) \nonumber
\\[0.21 cm]
& = & \frac{1}{E_0} ~ \int_0^1 dx_1~ \int_0^1 dx_2~
G_{Nj}(x_1)~F_{j\ell}(x_2)~\delta(x-x_1 \, x_2)\,,
\label{eq:scal_low}
\end{eqnarray}
with $x=E/E_0$. $G_{Nj}(x) ~ {\rm d}x$ is the average number of
particles $j$ created in the shower (in interactions of $N$
and secondary particles with air nuclei) 
with a fraction of energy in
the interval $[x,x+{\rm d}x]$. 
The function $G_{Nj} (x)$ is related to
$A_{Nj} (\alpha)$ by the expression:
\begin{equation}
A_{Nj} (\alpha) = \int_0^1 dx ~x^{\alpha -1} ~G_{N j} (x)\,.
\end{equation}
$G_{N j}(x)$ can therefore be calculated as the 
inverse Mellin transform of $A_{N j} (\alpha)$.

In the high-energy limit the lepton yield takes the form
\begin{eqnarray}
Y_{N \to \ell}^{(j)} (E, E_0, \theta) & = &
\frac{\varepsilon_j}{E_0^2 \; \cos \theta} ~ R_{N\ell}^{(j)} \left(
x \right)
= \frac{\varepsilon_j}{E_0^2 \; \cos \theta} ~ \left[
\frac{H_{Nj}}{x_1} \otimes F_{j\ell} \right] \left( x \right)
\nonumber
\\[0.21 cm]
& = & \frac{\varepsilon_j}{E_0^2 \; \cos \theta} ~\int_0^1 dx_1~
\int_0^1 dx_2~ \frac{H_{Nj}(x_1)}{x_1}~F_{j\ell}(x_2)~\delta(x-x_1
\, x_2)\,. \label{eq:scal_hig}
\end{eqnarray}
Again, $H_{Nj}(x)$ includes the production of hadrons $j$ in
primary and secondary interactions inside the shower started by
$N$. This function is related to 
$B_{Nj} (\alpha)$ by
\begin{equation}
B_{Nj} (\alpha) = \int_0^1 dx ~x^{\alpha -1} ~H_{N j} (x)
\end{equation}
and can therefore be calculated as its 
inverse Mellin transform.

It is straightforward to ckeck that for a power-law 
primary flux, using Eqs.~(\ref{eq:scal_low}) and (\ref{eq:scal_hig})
to perform the integration in Eq.~(\ref{eq:inty}),
one recovers for the lepton fluxes the results in 
Eqs.~(\ref{eq:form1}) and (\ref{eq:form2}).

\section{Atmospheric muons from unflavored mesons}
The decay of the unflavored mesons
\{$\eta$, $\eta^\prime$, $\rho^\circ$, $\omega$, $\phi$\} 
into final states that contain a $\mu^+ \mu^-$ pair also
contributes to the atmospheric muon fluxes.
It is straightforward to use the methods outlined in the previous section 
to estimate this contribution.

Using the assumptions (i), (ii) and (iii) the
muon flux from $\eta$ mesons is
\begin{equation}
\frac{\phi_\mu^{(\eta)} (E)}{\left ( K \; E^{-\alpha} \right )}
\simeq A_{\eta} (\alpha) ~Z_{\eta \mu} (\alpha)
=\left [
\frac{Z_{N\eta}}{1- Z_{NN}} +
\frac{Z_{N\pi} \; Z_{\pi\eta} }{(1- Z_{NN}) \; (1-Z_{\pi\pi}) }
\right ] ~ Z_{\eta \mu} (\alpha)\,,
\label{eq:feta}
\end{equation}
with similar expressions for all other unflavored mesons. 
Eq.~(\ref{eq:feta}) provides an estimate of 
the sum of the $\mu^+$ and $\mu^-$ fluxes, 
taking into account the production of eta mesons both
in nucleon and pion interactions
(we neglect the smaller contribution from kaon interactions).
The contributions from the decay of unflavored mesons to the 
positive and negative muon fluxes are identical.

The decay $Z$--factors are computed 
from the measured 
branching fractions into states with muon pairs and 
from the shape of the muon
energy spectra (in Fig.~\ref{fig:decay_spectra}). 
Numerical estimates for these factors are given in 
Table~\ref{tab:zetadec}.
These estimates have an uncertainty of order 20\%
associated to the experimental error in the measurement 
of the relevant branching ratios.
\begin{table} [b]
\begin{center}
\begin{tabular}{| c | c | c | }
\hline
$x$ & $Z_{x \mu} (2.7)$ ($ \times 10^{-4}$) &
$Z_{x \mu} (3.0)$ ($ \times 10^{-4}$) \\
\hline
$\eta$ & 1.37 & 1.12 \\
$ \eta^\prime$ & 0.43 & 0.35 \\
$\rho^\circ$ & 0.33 & 0.30 \\
$ \omega$ & 1.00 & 0.86 \\
$ \phi$ & 2.15 & 1.93 \\
\hline
\end{tabular}
\caption{Decay $Z$ factors for unflavored mesons.
\label{tab:zetadec}
}
\end{center}
\end{table}

The calculation of the 
$Z_{N\eta}$ and $Z_{\pi \eta}$ factors 
 obviously requires the modeling of unflavored meson 
production in nucleon and (less critically) pion 
interactions. 
The hadronic $Z$--factors entering Eq.~(\ref{eq:feta}) are
\begin{equation}
\begin{aligned}
& Z_{N \eta} = Z_{p \eta} = Z_{n \eta} \; ,
\\
& Z_{\pi \eta} = Z_{\pi^+ \eta} = Z_{\pi^- \eta} \; ,
\\
& Z_{NN} = Z_{pp} + Z_{pn} = Z_{nn} + Z_{np} \; ,
\\
& Z_{\pi\pi} 
= Z_{\pi^+ \pi^+} + Z_{\pi^+\pi^-} = Z_{\pi^- \pi^+} + Z_{\pi^-\pi^-} \,,
\end{aligned}
\end{equation}
where we have left the dependence on $\alpha$ implicit 
and have used isospin symmetry.

\begin{figure} [ht]
\begin{center}
\includegraphics[width=16.0cm]{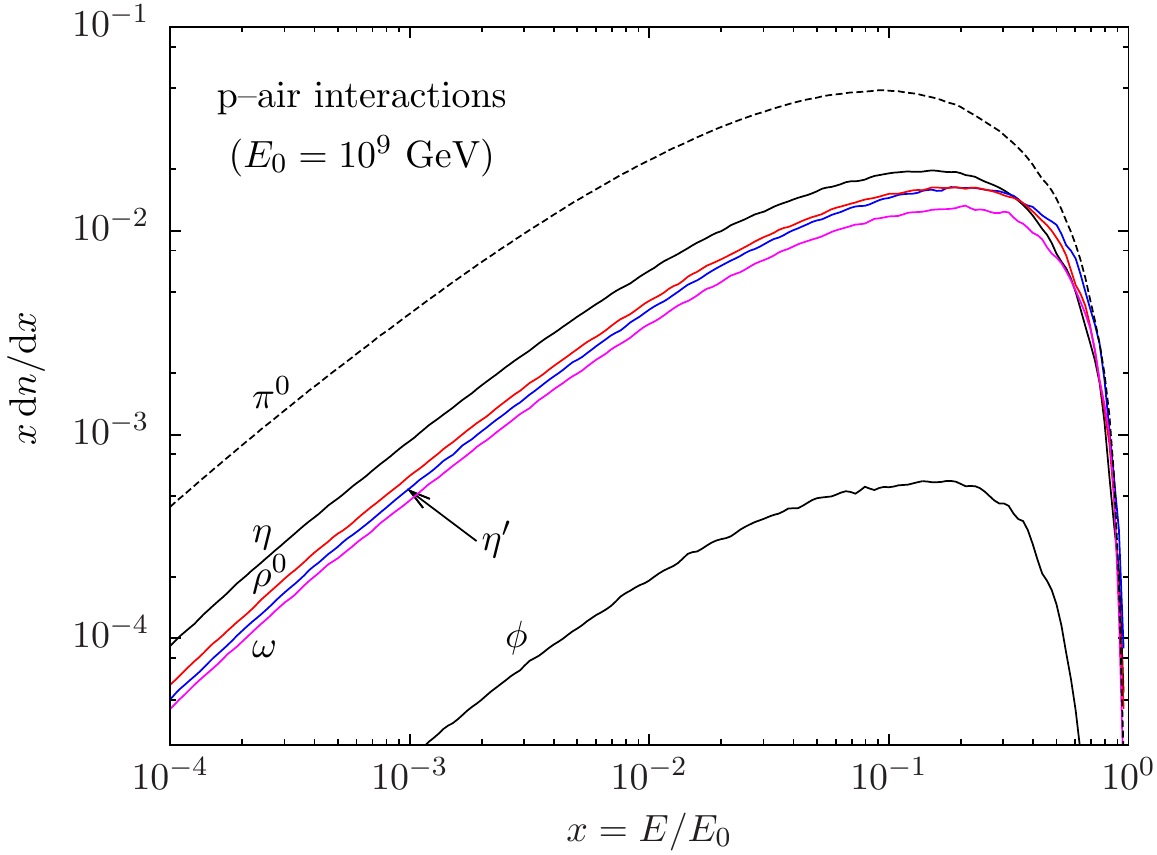}
\end{center}
\caption {\footnotesize
Inclusive spectra of flavorless mesons in $p$--air interactions
(for $E_0 = 10^{15}$~eV). The $\pi^\circ$ spectrum includes the
contribution of the decay of unstable resonances.
\label{fig:unflavored_spectra} 
}
\end{figure}

The inclusive meson spectra have been obtained 
from a Sibyll \cite{sibyll} Montecarlo simulation
(see Fig.~\ref{fig:unflavored_spectra}), and we have then 
evaluated the corresponding $Z$--factors 
through numerical integration. The results
for $\alpha = 2.7$ and
$\alpha = 3$ are listed in Table~\ref{tab:zhunflav}.
\begin{table} [hbt]
\begin{center}
\begin{tabular}{| c | c c | c c | c c | }
\hline
~~ 
& $Z_{p j}(2)$ & $Z_{\pi j}(2)$ 
& $Z_{p j}(2.7)$ & $Z_{\pi j}(2.7)$ 
& $Z_{p j}(3.0)$ & $Z_{\pi j}(3.0)$ \\ 
\hline
$\eta$ & 0.066 & 0.094 & 0.014 & 0.029 & 0.0087 & 0.021 \\
$\eta^\prime$ & 0.052 & 0.074 & 0.013 & 0.027 & 0.0086 & 0.020 \\
$\rho^\circ$ & 0.054 & 0.077 & 0.013 & 0.026 & 0.0082 & 0.019 \\
$\omega$ & 0.040 & 0.060 & 0.010 & 0.021 & 0.0066 & 0.016 \\
$\phi$ & 0.0019 & 0.0020 & 0.00038 & 0.00047 & 0.00022 & 0.00029 \\
All & 0.22 & 0.31 & 0.051 & 0.10 & 0.032 & 0.076 \\
\hline
\end{tabular}
\end{center}
\caption{$Z$ factors for the production of unflavored mesons in
proton and pion collisions  with an air nucleus.
The   inclusive  spectra are calculated  with the Sibyll Montecarlo  
code \protect\cite{sibyll}
\label{tab:zhunflav}
}
\end{table}
The combination
\begin{equation}
Z_{\rm unflav} (\alpha) = 
 \sum_{j \in \{\eta,\eta^\prime, \rho^\circ,\omega,\phi\}} Z_{N j} (\alpha) 
\label{eq:zunflav0}
\end{equation}
is also plotted as a function of $\alpha$ in Fig.~\ref{fig:zunflav}.

The contribution to the muon flux from unflavored mesons results 
from the addition 
\begin{equation}
\frac{\phi_\mu^{\rm unfl} (E) }{\left ( K \; E^{-\alpha} \right )}
= \sum_{j \in \{\eta,\eta^\prime, \rho^\circ,\omega,\phi\}} A_j (\alpha) \; 
Z_{j\ell} (\alpha) = C_\mu^{\rm unflav} (\alpha)\,.
\label{eq:f_unfl0}
\end{equation}
From the $Z$--factors in Tables~\ref{tab:zetadec}
and~\ref{tab:zhunflav} we obtain
\begin{equation}
\begin{aligned}
& C_\mu^{\rm unflav} (\alpha = 2.7) \simeq 6.2 \times 10^{-6} \; ,
\\[0.21 cm]
& C_\mu^{\rm unflav} (\alpha = 3.0) \simeq 3.1 \times 10^{-6} \; .
\end{aligned}
\label{eq:num_unflavored}
\end{equation}

\begin{figure} [ht]
\begin{center}
\includegraphics[width=16.0cm]{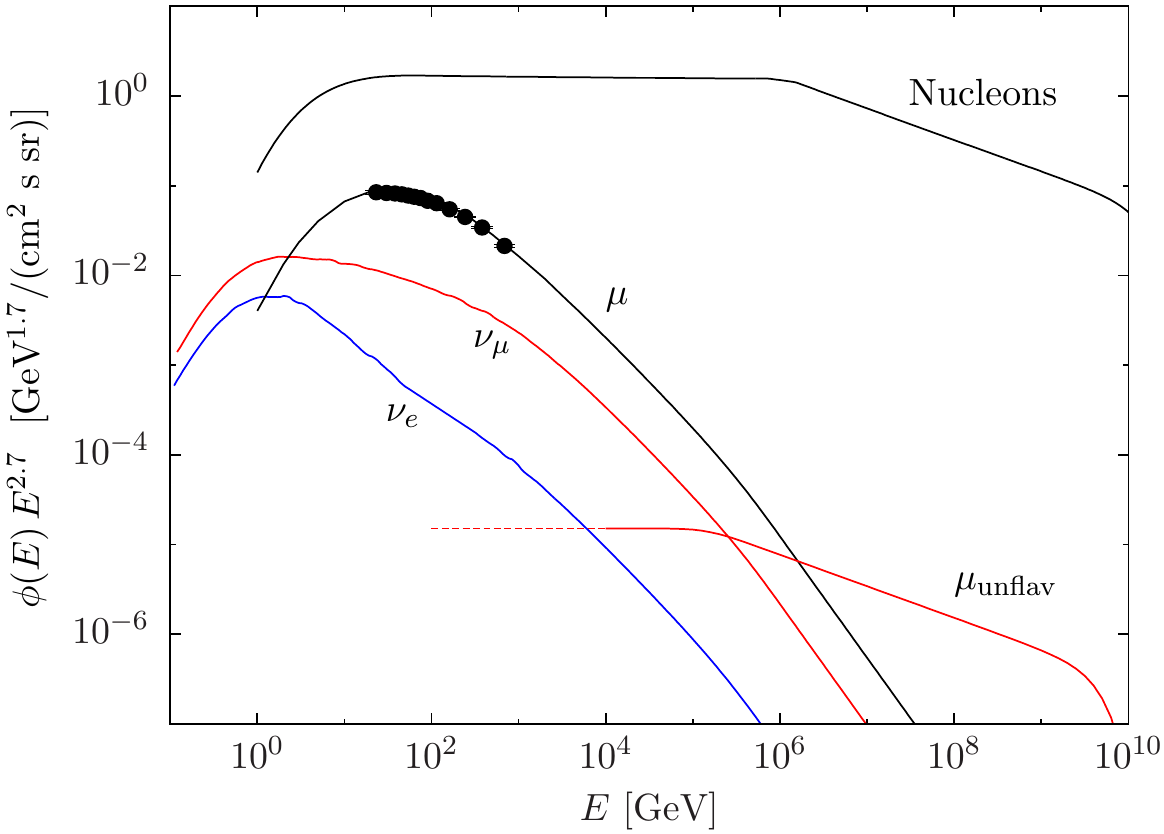}
\end{center}
\caption {\footnotesize
The top line shows the fit of the nucleon flux
used in 
\protect\cite{Barr:2004br}. 
The curves labeled as $\mu$, $\nu_\mu$ and
$\nu_e$ indicate the vertical fluxes of
atmospheric $\mu^\pm$, 
$\nu_\mu + \overline{\nu}_\mu$ and
$\nu_e + \overline{\nu}_e$. 
The calculation of the fluxes extends to higher energy of
the results of \protect\cite{Barr:2004br} 
and coincides with
those results for $E \lesssim 30$~TeV.
The contribution of the decay of unflavored mesons to the muon
flux is shown as the thick (red) curve.
The points are the measurement of the muon flux of the
L3 detector \protect\cite{L3-muons}.
\label{fig:lepton_spectra} }
\end{figure}

In the regions of the spectrum where the 
primary nucleon flux is not well described 
by a single power law
it is possible to compute $\phi_\mu^{\rm unfl} (E)$ 
from the muon yields. 
The results of such calculation that we obtain 
using the Sibyll Montecarlo code are given in 
Fig.~\ref{fig:lepton_spectra}. 
The top line there shows the all-nucleon
primary flux. The thick red line is our
estimate of the muon flux from the
electromagnetic decay of unflavored mesons.
We include the conventional
$\mu^+ + \mu^-$,
$\nu_\mu + \overline{\nu}_\mu$
and
$\nu_e + \overline{\nu}_e$ fluxes
(from the vertical direction)
 taking into account only the decay of
charged pions and kaons.
The data points are from the L3 detector \cite{L3-muons}.

Inspection of Fig.~\ref{fig:lepton_spectra} shows that 
for the vertical direction
($\theta = 0$) the contribution of unflavored
muons overtakes the standard contribution from pion and kaon decay 
at $E_\mu \simeq 1.6 \times 10^6$~GeV.
This result can be also estimated combining 
Eqs.~(\ref{eq:num_standard}) and (\ref{eq:num_unflavored}).
The unflavored contribution is isotropic, 
while the standard contribution
grows with increasing zenith angle 
proportionally to $F_{\rm zenith} (\theta)$.

\subsection{Unflavored meson production}
We would like to analyze the uncertainties in the
modeling of unflavored-meson production in hadronic interactions,
and in this subsection we give a qualitative discussion. 

The most important quantities for the prediction of the
muon flux are the fraction of the projectile energy 
carried by the unflavored mesons produced in the collision 
and the shape of their energy spectra.
To estimate the energy fraction 
$\langle x_j \rangle = \langle E_j\rangle/E_0$ carried by
the meson type $j$,
one can make some simple considerations.
The initial energy $E_0$ of the projectile
is divided among three classes of particles: 
baryons, antibaryons and mesons,
\begin{equation}
E_0 \simeq E_{qqq} + E_{\overline{q}\overline{q}\overline{q}} +
E_{q\overline{q}} \,.
\end{equation}
The energy $E_{q\overline{q}}$ carried by mesons 
is in turn subdivided
among different particle types:
\begin{equation}
 E_{q\overline{q}} = \sum_{j \in \{ {\rm mesons} \}} E_j
\end{equation}
The summation is over the {\it primary} 
mesons (before the decay of unstable particles).
It is a good approximation to neglect heavy quarks 
and include in the sum only the 
18 mesons that compose the scalar and vector nonets of $SU(3)$.
Meson production can be modeled as a
two-step process: in the first step $q_i \overline{q}_i$ pairs are
created and recombined between each other and with the valence
quarks of the interacting nucleons;
in the second step the states $q_j\overline{q}_k$ are
projected into physical mesons, whose quark content is known.
The scalar unflavored mesons have the quark content
\begin{equation}
\begin{aligned}
\pi^\circ & = \frac{1}{\sqrt{2}} \, \left (
u \overline{u} -
d \overline{d}
\right )\,; \\
\eta & = \frac{1}{2} \, \left ( u \overline{u} + d \overline{d}
\right )
- \frac{1}{\sqrt{2}} \, s \overline{s} \,; \\
\eta^\prime & = \frac{1}{2} \, \left ( u \overline{u} + d
\overline{d} \right )
+ \frac{1}{\sqrt{2}} \, s \overline{s}\;, \\
\end{aligned}
\end{equation}
whereas for the vector mesons:
\begin{equation}
\begin{aligned}
\rho^\circ & = \frac{1}{\sqrt{2}} \, \left (
u \overline{u} -
d \overline{d}
\right ) \,;\\
\omega & = \frac{1}{\sqrt{2}} \, \left ( u \overline{u} + d
\overline{d}
\right ) \,;\\
\phi & = s \overline{s} \,. \\
\end{aligned}
\end{equation}

Assuming that each light
flavor combination $q_j\overline{q}_k$
has on average the same energy
and neglecting meson mass differences,
the fraction of energy carried by the different species
can be calculated in terms of just two parameters:
the probability $P_s$ of producing
an $s\overline{s}$ pair (with $P_u = P_d = (1-P_s)/2$),
and the probability $P_{\rm scalar}$
to project the state $q_j\overline{q}_k$ into
a spin--0 meson.
The energy fractions can then be estimated as
\begin{equation}
\begin{aligned}
 & \frac{\langle E_{\eta} \rangle} {E_{q\overline{q}} }
\simeq \frac{\langle E_{\eta^\prime} \rangle} {E_{q\overline{q}} }
\simeq P_{\rm scalar} \;\left [
 \; \frac{(1- P_s)^2}{8} +
\frac{P_s^2}{2}
\right ] \\
& \frac{\langle E_{\rho^\circ} \rangle} {E_{q\overline{q}} }
\simeq \frac{\langle E_{\omega} \rangle} { E_{q\overline{q}}}
\simeq (1-P_{\rm scalar})
 \; \frac{(1- P_s)^2}{4}
\\
& \frac{\langle E_{\phi} \rangle} {E_{q\overline{q}}} \simeq
(1-P_{\rm scalar}) \; P_s^2\,.
\end{aligned}
\end{equation}
For completeness, the average energy taken by the other mesons is
\begin{equation}
\begin{aligned}
 &
\frac{\langle E_{\pi^\circ} \rangle} {E_{q\overline{q}}} \simeq
\frac{\langle E_{\pi^+} \rangle} {E_{q\overline{q}}} \simeq
\frac{\langle E_{\pi^-} \rangle} {E_{q\overline{q}}} \simeq P_{\rm
scalar}
 \; \frac{(1- P_s)^2}{4} \\
& \frac{\langle E_{K^+} \rangle} {E_{q\overline{q}}} \simeq
\frac{\langle E_{K^-} \rangle} {E_{q\overline{q}}} \simeq
\frac{\langle E_{K_L} \rangle} {E_{q\overline{q}}} \simeq
\frac{\langle E_{K_S} \rangle} {E_{q\overline{q}}} \simeq P_{\rm
scalar}
 \; \frac{(1- P_s) \, P_s}{2}\,.
\end{aligned}
\label{eq:mes1}
\end{equation}
For the corresponding vector particles
($\rho$ and $K^*$)
one can use Eq.~(\ref{eq:mes1}) with the substitution
$P_{\rm scalar} \to 1-P_{\rm scalar}$. 

For a numerical estimate we can use $E_{q\overline{q}}/E_0 \simeq 0.6$ 
(with most of the remaining energy carried by one leading baryon) 
and, following Field and Feynman 
\cite{Field:1977fa}
and the Lund fragmentation algorithm \cite{pythia}, 
$P_{\rm scalar} \simeq 0.5$ and $P_s \simeq 0.13$. 
With these assumptions the 5 unflavored mesons carry together 
an energy fraction
 $\langle E_{\rm unflav} \rangle /E_0 \simeq 0.18$.
This estimate 
depends only weakly 
on the values chosen for $P_{\rm scalar}$ and $P_s$.
The minimum value, 
$\langle E_{\rm unflav} \rangle \simeq 0.13$, 
is obtained 
for $P_{\rm scalar} = 1$ and $P_s = 0$. 

It is reasonable to expect that the energy spectrum of 
unflavored mesons 
is similar or slightly harder than the spectrum 
observed for pions (which may come from 
a longer decay chain of unstable primary mesons). 
A simple 2--parameter form 
for the inclusive energy spectrum $F_{N j} (x)$ is:
\begin{equation}
F_{N j} (x) =
\langle x_j \rangle \, (1+n_j)
\;\frac{(1-x)^{n_j}}{x}\,.
\label{eq:forma}
\end{equation}
where $\langle x_j \rangle$ is the energy fraction carried by particle $j$
and $n_j$ a shape parameter in the range 3--4.
The $Z_{Nj}$ moments corresponding to (\ref{eq:forma}) are
\begin{equation}
Z_{Nj} (\alpha) = \langle x_j \rangle ~
 \frac{(1+n_j) \, \Gamma(\alpha -1) \; \Gamma (n_j+2)}
{\Gamma (n_j + \alpha)}\,.
\label{eq:zex}
\end{equation}
Note that 
 $Z_{kj} (2) = \langle x_j \rangle$
and $Z_{kj} (3) = \langle x_j \rangle/(n_j+2)$.
With these simple considerations it is straightforward to 
obtain results that are 
 close to those obtained with the Sibyll Montecarlo code.

\subsection{Comparison with Pythia}
\begin{figure} [ht]
\begin{center}
\includegraphics[width=16.0cm]{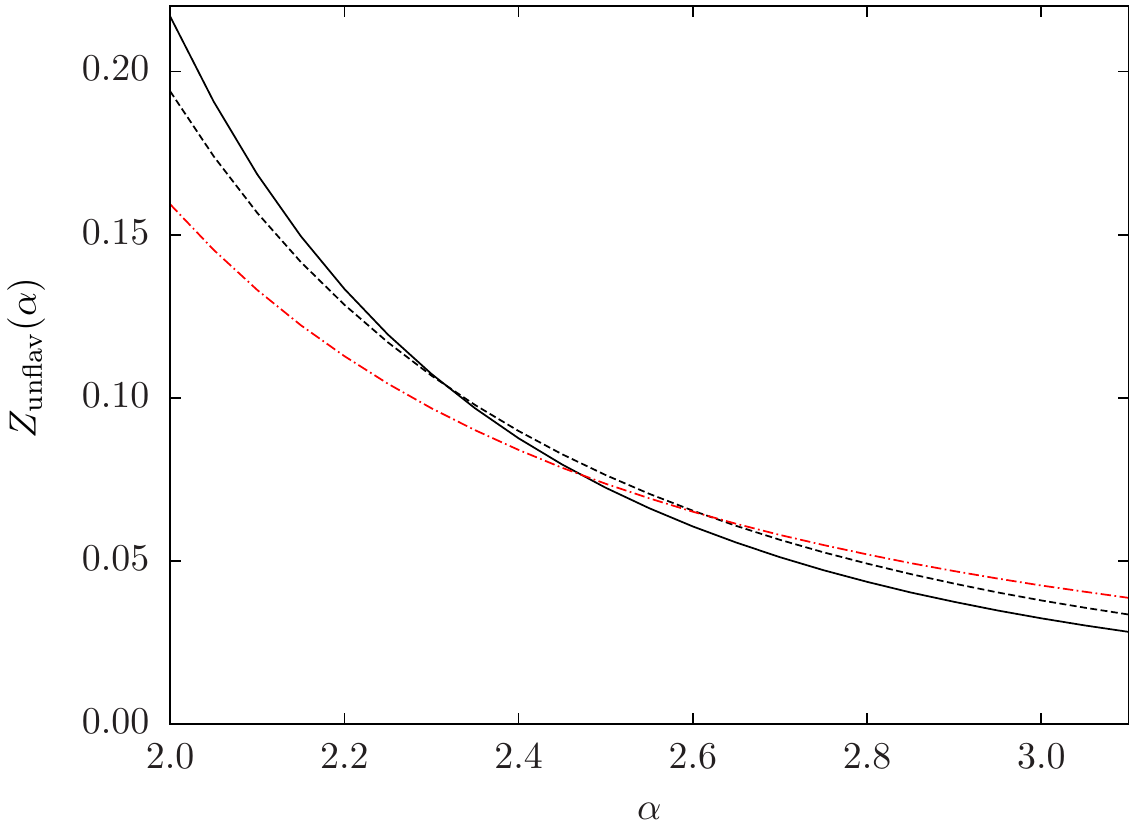}
\end{center}
\caption {\footnotesize
Plot of the quantity $Z_{\rm unflav}(\alpha)$ 
(defined  in  equation (\protect\ref{eq:zunflav0})) 
as a function of $\alpha$.
The three curves correspond to calculations performed
with the Sibyll montecarlo code for $pp$ 
(dashed line) and $p$--air interactions (solid line),
and with the Pythia code (dot--dashed line) for $pp$ interactions.
\label{fig:zunflav} }
\end{figure}

To estimate the systematic uncertainty 
associated to our calculation 
we have performed a second calculation 
of the inclusive spectra of unflavored mesons
using the Pythia Montecarlo code.
Since Pythia does not support collisions with a nucleus, we have 
simulated $pp$ collisions at $10^6$ GeV, and compared 
the results with the ones from 
Sibyll for the same type of interactions.
The values obtained for the combination $Z_{\rm unflav} (\alpha)$ 
defined in Eq.~(\ref{eq:zunflav0}) are shown in 
Fig.~\ref{fig:zunflav}.
It is apparent that unflavored meson production in 
$pp$ interactions is qualititively similar in 
the Pythia and Sibyll codes.
Unflavored mesons carry approximately a fraction 0.19 of the 
projectile energy if produced with the Sibyll code; 
for the Pythia simulation this energy fraction is reduced
to 0.16 (a 20\% difference). 
The energy spectrum in Pythia is however slightly harder.
Accordingly, with growing $\alpha$, the Pythia $Z$--factor
decrease a little more slowly, and the difference 
between the models is reduced.
For $\alpha \simeq 2.6$ the $Z$--factors 
for unflavored meson production
calculated with the 2 codes coincide.
For $\alpha \simeq 3$ the Pythia code gives a 
result 10\% larger.
The two codes also show some differences in the 
relative importance
of the different mesons. In Pythia the scalar (vector) 
mesons are less (more) important
with respect to Sibyll.

In summary, the description of unflavored meson production
in Sibyll and Pythia in $pp$ interactions agrees at the level of 10--15\%.
This level of agreement is however likely to be an
underestimate of the theoretical uncertainties, because 
the two Montecarlo codes use very similar assumptions.

The calculation of cosmic ray showers requires also
the description of hadronic interactions with a nuclear target.
The Sibyll code allows to compute
the $Z$ factors for both $pp$ and $p$--air interactions.
The differences are small but not negligible.
For a nuclear target the unflavored 
mesons carry a slightly larger fraction of the energy,
but have a softer spectrum (see fig.~\ref{fig:zunflav}). 
The first effect is a consequence of the fact 
that in nuclear interactions the 
leading baryon  is  less energetic than in $pp$ scattering,
and therefore more energy goes into  mesons  production,
on  the other  hand  for  a nuclear targer  all inclusive  
spectra  are  softer.

\section{Leptons from charm decay}
Weakly decaying charmed particles ($D^\circ$, $D^+$, $D_s^+$,
$\Lambda_c$ and their antiparticles) have a significant
probability to decay in semileptonic modes such as $D^\circ \to
K^- \mu^+ \nu_\mu$ or $D^\circ \to K^- e^+ \nu_e$, and therefore
are sources of atmospheric muons and neutrinos. 
The production of charmed particles, however, 
is dynamically suppressed with respect to 
the production of pions and
kaons, and their contribution to the lepton fluxes is 
usually negligible and remains undetected.
On the other hand, charmed particles have a lifetime of order 
$\tau \sim 10^{-12}$~seconds,
and decay with probability close to one up to very high energy.
The critical energies $\varepsilon_j$ for 
$D^\circ$, $D^\pm$, $D_s^\pm$, $\Lambda_c$
 are (0.38, 0.96, 0.85, 2.4)$\times 10^8$~GeV, several orders 
of magnitude larger than the ones for pions and kaons.
Therefore, as the energy grows pion and kaon decay 
is suppressed and the {\it prompt} contribution from
charm decay will necessarily overtake the standard lepton fluxes.

The estimate of the lepton fluxes from charm 
decay has been the subject of many studies 
\cite{Gondolo:1995fq,Pasquali:1998ji,Costa:2000jw,Martin:2003us,Enberg:2008te},
with results that span a very broad range.
In most cases the prediction for this contribution to 
the muon flux 
remains always below the one from 
unflavored meson decay discussed in the previous section.
If this were the case the charm contribution would only be
observable in measurements of neutrino fluxes.
We do not intend to perform here a new calculation of the 
atmospheric lepton fluxes from charm decay nor 
a critical review of existing predictions. However,
we would like 
to discuss under which conditions this muon flux is 
above the expected one from unflavored mesons.

Very likely the production of charmed particles does not 
obey a scaling law of the type in Eq.~(\ref{eq:scal0}). 
A first order estimate of this contribution  
can however still be  obtained  using 
the simple analytic  expressions  discussed  before,
treating the hadronic $Z$--factors  as energy-dependent quantities.
For $E_\ell < 10^7$~GeV ({\it i.e.,} a parent charmed 
particle that decays with probability close to 1) and 
approximating the primary nucleon flux as a power law, 
the lepton flux from charm decay 
can be estimated as
\begin{equation}
\frac{ \phi_{\ell}^{c\overline{c}} (E) }
{\left ( K ~E^{-\alpha} \right ) }
\simeq C_\ell^{c\overline{c}} (\alpha,E)
 \simeq
\sum_{j \in \{ D_0, \overline{D}_0, D_s^\pm, \Lambda_c \}}
A_j(\alpha,E) ~Z_{j \ell} (\alpha)\,. 
\label{eq:charm0}
\end{equation}
Including the production of charmed particles in nucleon 
and pion collisions, the factors
$A_j(\alpha,E)$ are:
\begin{equation}
A_{j} (\alpha,E)
=
 \frac{Z_{N j} (\alpha,E)}{1- Z_{NN} (\alpha)}
+
 \frac{Z_{N \pi} (\alpha) \; Z_{\pi j} (\alpha,E)}
{[1- Z_{NN} (\alpha)] \, [1- Z_{\pi\pi} (\alpha)]}\,,
\label{eq:acharm}
\end{equation}
where for simplicity we have assumed equal cross sections 
for charm production 
in $p/n$ or $\pi^\pm$ interactions with an air nucleus. 

The contribution from charm produced in pion interactions is 
suppressed by a factor $Z_{N \pi}$, and is expected to introduce just
a 20--30\% correction.
Therefore, for a  first order estimate it is sufficient to model 
charm production in nucleon interactions.
To describe the production of the charmed hadron 
type $j$ in the forward
hemisphere we follow the suggestion in \cite{Frixione:1997ma} and 
parametrize the inclusive spectrum $F_{Nj} (x, E_0)$ as
\begin{equation}
F_{Nj} (x, E_0) =
\frac{A \, \sigma^{pp}_{c\overline{c}} (E_0)}
{\sigma^{pA}_{\rm inel} (E_0) }
\; p_j \; (n_j+1) (1-x)^{n_j}\,,
\label{eq:charm_param1}
\end{equation}
This expression can be integrated in the entire interval $x\in [0,1]$. 
The corresponding total charm cross section is
\begin{equation}
\sigma_{c\overline{c}}^{pA} (E_0) \simeq A 
~\sigma_{c\overline{c}}^{pp} (E_0) \,,
\end{equation}
which scales linearly with the mass number $A$ of the target nucleus. 
The quantity $p_j$ is the fraction of charm events 
that contain the species $j$, with $\sum_j p_j = 1$. 
The $Z$--factor that corresponds to (\ref{eq:charm_param1}) is
\begin{equation}
Z_{Nj} (\alpha, E) \simeq
\frac{A \, \sigma_{c\overline{c}}^{pp} (E)}
{\sigma_{pA}^{\rm inel} (E) } ~p_j
~\hat{z} (\alpha, n_j) 
= 
\frac{A \, \sigma_{c\overline{c}}^{pp} (E)}
{\sigma_{pA}^{\rm inel} (E) } ~p_j
~ \frac{\Gamma (\alpha) \; \Gamma (n_j + 2)}
{\Gamma (n_j + \alpha + 1) } \;. 
\label{eq:zcharm}
\end{equation}
Note that 
$\hat{z}(1,n) = 1$, 
$\hat{z}(2,n) = 1/(n+2)$, and
$\hat{z}(3,n) = 2/[(n+2)(n+3)]$.

The modeling of the production of leading charmed baryons 
remains an important open problem.
This process results into a final state with a $\Lambda_c$ 
(the longest lived charmed baryon)
and, after its decay, high-energy leptons. 
For this reason we decompose the cross section 
in 2 parts, 
\begin{equation}
\sigma_{c\overline{c}} =
\sigma_{\Lambda_{c} \overline{D}}
+
\sigma_{D \overline{D}} \;,
\end{equation}
where the first term accounts for the production of 
charmed baryons by non perturbative processes.
It is natural to expect the energy spectrum of the 
$\Lambda_c$ to be significantly harder 
than the spectrum of $D$'s. Choosing as reference point
$n_D \simeq 5$ and $n_{\Lambda_c} \simeq 1$, we obtain
\begin{equation}
\begin{aligned}
C_\mu^{c\overline{c}} (\alpha = 3) \simeq
&
1.2 \times 10^{-6} \;
\left [\frac{\sigma_{D\overline{D}}^{pp}}{100~\mu{\rm barn}} \right ]
~\frac{\hat{z}(3, n_D)}{\hat{z}(3,5)} +
 \\[0.21 cm]
+ &
 1.5 \times 10^{-6} \;
\left [
\frac{\sigma_{\Lambda_c\overline{D}}^{pp} }{100~\mu{\rm barn}}
 \right ]
~\left (
 0.6 \; \frac{\hat{z}(3, n_{\Lambda_c})}{\hat{z}(3,1)}
 + 0.4 \; \frac{\hat{z}(3, n_D)}{\hat{z}(3,5)}
\right )
\end{aligned}
\label{eq:c-num}
\end{equation}
This prediction assumes isospin symmetry, 
that (primary) scalar and vector charmed mesons are produced with 
the ratio 1/3, and that the production
of $D_s$ is suppressed by a factor $\simeq 0.12$ with 
respect to the production of charmed 
particles with zero strangeness.

Comparing Eqs.~(\ref{eq:c-num}) and (\ref{eq:num_unflavored}) 
one can see that the
charm contribution to the atmospheric muon flux is below the 
contribution from
unflavored mesons unless the cross section 
(at $E_0 \simeq 10^6$~GeV) is 
larger than $\sigma_{c\overline{c}}^{pp} \simeq 100~\mu$barn 
or the energy spectrum 
of charmed particles is surprisingly hard. 

The neutrino spectra generated by charmed particle decay can be estimated
including the appropriate energy spectrum of the neutrinos.
Because of the ($V-A$) properties of the matrix element 
these spectra are 
a little harder than the corresponding muon spectrum. The decay $Z$--factors 
(for $\alpha = 3$) are in the ratios:
\begin{equation}
\begin{aligned}
&
Z_{D \nu_\mu} (3) \simeq
Z_{D \nu_e} (3)
 \simeq 1.25 ~ Z_{D \mu} (3)
\\[0.21cm]
&
Z_{\Lambda_c \nu_\mu} (3)
\simeq
Z_{\Lambda_c \nu_e} (3)
\simeq 1.16 ~ Z_{\Lambda_c \mu} (3)
\end{aligned}
\label{eq:zc1}
\end{equation}
The different $\nu/\mu$ ratios 
for $D$ and $\Lambda_c$ decay are a consequence of
the difference in the available phase space in the two cases.
In conclusion one expects that, 
without the inclusion of neutrino oscillations, 
 the $\nu_e$ and $\nu_\mu$ spectra from charm decay 
are approximately 20\% higher that the 
corresponding muon flux. This can be considered a 
robust prediction, in the sense that it is essentially
independent of the modelling used for charm production.

The flux of $\nu_\tau$ generated by the chain decay of $D_s^\pm$ 
is approximately 30 times smaller than the $\nu_e$ or $\nu_\mu$
fluxes. This estimate is obtained assuming 
$\sigma_{D_s}/\sigma_{D} \simeq 0.12$ 
and taking into account the relevant decay branching ratios and 
energy spectra. 

The decay probability of charmed particles 
becomes less than unity for
energies larger than $\sim 10^{7}$~GeV, and therefore 
the expression in Eq.~(\ref{eq:charm0}) 
becomes a poor approximation. 
To describe the resulting lepton fluxes at such energies 
it is necessary to introduce their critical energies
$\varepsilon_j$ and use a functional form of the type given in
Eq.~(\ref{eq:form3}).

\section{Photon conversion into $\mu^+\mu^-$ pairs}
A potentially interesting source of atmospheric muons is the 
photon conversion into a pair of muons:
\begin{equation}
\gamma + Z \to \mu^+\mu^- + Z
\label{eq:muon_pairs} 
\end{equation}
where $Z$ is the electric charge of an air nucleus.
This process is suppressed with respect to the production 
of $e^+e^-$ pairs by 
a factor of $(m_e/m_\mu)^2 \simeq 2.3 \times 10^{-5}$. 
However, high energy showers
contain a very large number of photons, 
and it is not immediately obvious that this contribution 
to the muon flux is entirely negligible.

To estimate the contribution of photon conversions into $\mu^+\mu^-$ pairs 
to the atmospheric muon flux it is possible to use the 
same methods discussed above.
The development of an electromagnetic shower at 
high energy is controled by the 
processes of bremsstrahlung and pair production, 
which are described by the 
scaling functions
\begin{equation}
\frac{d\sigma_{e\to e \gamma}}{dv} = X_0 \; \frac{A}{N_A} \;
\varphi(v)\,, 
\label{eq:brems}
\end{equation}
with $v = E_{\gamma}/E_e$, and
\begin{equation}
\frac{d\sigma_{\gamma \to e e}}{du} = X_0 \; \frac{A}{N_A} \;
\psi(u)\,, 
\label{eq:pair1}
\end{equation}
with $u = E_{e^+}/E_{\gamma}$. 
The well known expressions for $\varphi(v)$ and $\psi(u)$ can be 
found, for example, in \cite{Rossi-Greisen}.
In a first approximation the production of muon pairs can be obtained
simply rescaling Eq.~(\ref{eq:pair1}):
\begin{equation}
\frac{d\sigma_{\gamma \to \mu \mu}}{du} \simeq
\left (\frac{m_e}{m_\mu} \right )^2 ~\frac{d\sigma_{\gamma \to e e}}{du}  \;.
\label{eq:pair2}
\end{equation}

Assuming a power law for the primary nucleon flux and scaling for the
hadronic interactions, it is straightforward to obtain that the resulting
contribution to the muon flux has the form:
\begin{equation}
\frac{\phi_{\mu}^{(\gamma)}(E)}{\left ( K \;E^{-\alpha} \right )}
\simeq C_\mu^{(\gamma)} (\alpha) \,.
\end{equation}
The $\alpha$--dependent constant $C_{\mu}^{(\gamma)} (\alpha)$ can be 
calculated as
\begin{equation}
\begin{aligned}
& C_{\mu}^{(\gamma)} (\alpha) \simeq 
\left [ \frac{Z_{N \gamma}}
{1-Z_{NN}} + \frac {Z_{p\pi} \, Z_{\pi\gamma}}
{(1-Z_{NN})(1-Z_{N\pi}} \right ] \times 
~~~~~~~~~~~~~~~~~~~~~~~~~~~~~~~~~~~~~~~~~
\\[0.2 cm]
& ~~ ~~~~~~~~~~~~~~~~~~~~~~~~~~~ 
\left [ \frac{A(\alpha-1) \, \sigma_\gamma } 
{A(\alpha-1) \, \sigma_\gamma - B(\alpha-1) \, C(\alpha-1)}
 \right ]
~ B(\alpha-1) ~\left ( \frac{m_e}{m_\mu} \right )^2 \,.
\end{aligned} 
\label{eq:gamma_mu}
\end{equation}
In this expression one can recognize the effects of an hadronic shower
that is the source of an electromagnetic one.
In Eq.~(\ref{eq:gamma_mu}) we have left implicit the 
 $\alpha$ dependence of the 
hadronic $Z$ factors $Z_{N \gamma}$ and $Z_{\pi \gamma}$, the 
moments of the inclusive photon energy spectra in nucleon
and charged pion interactions. The photon spectrum 
is generated by the decay of neutral pions,
with smaller contributions from the decay of $\eta$ mesons
and other hadronic resonances.
The quantity $\sigma_\gamma$ and the functions 
$A(s)$, $B(s)$ and $C(s)$ were introduced by
Rossi and Greisen \cite{Rossi-Greisen}:
\begin{equation}
\sigma_\gamma = \int_0^1 ~du~ \psi(u)
\end{equation}
and
\begin{eqnarray}
A(s) & = & \int_0^1 ~dv ~\varphi(v)~ \left [
1- (1-v)^s \right ]
\\
B(s) & = & 
 2 \;\int_0^1 ~du~u^s ~\psi(u)
\\
C(s) & = &
 \int_0^1 ~dv~v^s ~\varphi(v) 
\end{eqnarray}

Using in Eq.~(\ref{eq:gamma_mu}) the numerical hadronic $Z$-factors 
obtained with Sibyll at $E \simeq 10^{6}$~GeV we obtain
\begin{equation}
C_\mu^{(\gamma)}
(\alpha = 3.0) \simeq 0.39 \times 10^{-6} \; .
\label{eq:num_gamma}
\end{equation}
For $\alpha \simeq 2.7$ the result is approximately 2.5 times larger.

Comparing this result with Eq.~(\ref{eq:num_unflavored}) one can 
see that this contribution to the atmospheric muon flux
is approximately one order of magnitude smaller than our 
estimate from unflavored meson decay.
Most predictions of the muon flux from charm decay 
are also above the result in Eq.~(\ref{eq:num_gamma}).
Photon conversion into muon pairs is therefore likely to
contribute only a small fraction of the atmospheric muon flux even at the 
highest energies.

\section{Summary and Discussion}
In this work we have discussed the main contributions to the atmospheric
lepton fluxes at very high energy. 
The standard contribution, due to the decay of charged pions and kaons, 
is suppressed at high energy because most of these 
long--lived particles interact, and only a small fraction
(inversely proportional to the energy) decays.
This suppression results in very steep energy spectra 
for the lepton fluxes,
with a slope approximately one unit larger than the one of the primary 
cosmic ray flux.
The zenith angle distributions of these lepton fluxes 
have also a characteristic shape, that for $\theta$ not
too large has the well known $(\cos \theta)^{-1}$ form.
Summing over particles and anti--particles 
(and neglecting neutrino oscillations) 
the lepton fluxes from pion and kaon decay are in the ratios:
\begin{equation}
\mu \div \nu_\mu \div \nu_e \div \nu_\tau \simeq 
1 \div 0.35 \div 0.02 \div 0
\end{equation}
These ratios reflect the fact that charged pions 
generate only $(\mu\nu_\mu)$ pairs, with the muon taking 
a larger fraction of the parent particle energy. The 
flux ratios have an uncertainty significantly smaller
than  the absolute value of the fluxes.
The remaining uncertainty is dominated  by the error
in the estimate of the  relative importance of kaon and  pion production,
and  is of order of  15\% (20\%)  for   $\mu/\nu_\mu$  ($\nu_e/\nu_\mu$).

The contributions to the lepton fluxes from charm decay 
are characterized by a zenith angle distribution that is isotropic 
for energies below $10^{7}$ GeV, and an energy spectrum
that roughly follows  the primary nucleon spectrum,
with corrections related to the energy dependence  of the charm production
cross sections.
 The prediction has however a large uncertainty
due to our poor understanding of the dynamics of charmed hadron production.
The ratios between the lepton fluxes
can be predicted with a much smaller uncertainty 
\begin{equation}
\mu \div \nu_\mu \div \nu_e \div \nu_\tau \simeq 
1 \div 1.2 \div 1.2 \div 0.04
\end{equation}

The decay of unflavored mesons generates an isotropic 
muon flux, that follows 
closely the shape of the primary cosmic ray spectrum.
Montecarlo codes like Sibyll of Pythia predict 
that this
contribution will overtake the standard contribution 
(for the vertical direction) at an energy 
$E \simeq 1.5 \times 10^{6}$~GeV.
Such result reflects that in these Montecarlo codes 
all mesons that compose the scalar and vector SU(3) nonets are 
produced with probabilites of the same order
(taking into account a suppression for strange quarks). 
This implies an abundant production
of $\eta$, $\eta^\prime$, $\rho^\circ$ and $\omega$
mesons, and using
the relevant branching ratios one can easily estimate
the atmospheric muon flux.
The prediction  can therefore be considered as quite robust.
It should be stressed that this flux is larger than most 
predictions for the charm contribution, 
and therefore it may be important for future observations.

We have also considered the flux generated by 
photon conversion into muon pairs.
The muon flux from this source is isotropic 
and has approximately the same energy spectrum 
as the flux from the decay of unflavored mesons,
but it is approximately 10 times smaller.
Photon conversion into muon pairs is therefore likely 
to be of little phenomenological importance.

It should be stressed that a large uncertainty 
in the prediction of the lepton fluxes at very 
high energy arises from our poor knowledge of the 
primary cosmic ray flux.
The relevant energy range is above the cosmic ray knee
at $E \simeq 3 \times 10^{6}$~GeV, 
where the observations have large errors. Still more
important is the fact that the composition of the 
cosmic ray flux in this energy range is 
very poorly known. 
The prediction of the lepton fluxes depends on the 
so called nucleon spectrum, 
\begin{equation}
\phi_0 (E_0) = \sum_A A^2 \; \phi_A (E_0 \, A)\,,
\label{eq:nucleon_flux}
\end{equation}
where $\phi_A(E)$ is the flux for the nuclear species of mass number $A$ and 
$E_0$ is the energy per nucleon. In Eq.~(\ref{eq:nucleon_flux})
the contribution of each nuclear species is weighted by a factor $A$ 
to account 
for the nucleon multiplicity, and a second factor $A$ is in 
the jacobian 
for the transformation from total energy to energy per nucleon.
If the cosmic ray flux is a power law of exponent $\alpha \simeq 3$ the
nucleon flux scales with composition 
$\propto \langle A^{2-\alpha} \rangle \simeq \langle A^{-1} \rangle$. 
A heavy composition corresponds then to a smaller nucleon flux, and to 
smaller lepton fluxes.

For the normalization of the nucleon flux shown in 
Fig.~\ref{fig:lepton_spectra}
the flux of muons from $\pi/K$ and  unflavored mesons decay  predicted here is of order
\begin{equation}
\left \langle \Phi_\mu^{\rm stand} (E_{\rm min} ) 
\right \rangle \simeq 
400 ~\left [\frac{\rm 10^{6}~{\rm GeV}}{E_{\rm min} } \right ]^{-3}  
~({\rm Km}^2{\rm yr}\, {\rm sr} )^{-1} \;,
\end{equation}
\begin{equation}
\Phi_\mu^{\rm unflav} (E_{\rm min} ) \simeq 
90 ~\left [\frac{\rm 10^{6}~{\rm GeV}}{E_{\rm min} } \right ]^{-2} 
~({\rm Km}^2{\rm yr}\, {\rm sr} )^{-1}
\end{equation}
(note the difference in the energy dependence).
For  the standard  contribution we have  performed an average over 
the entire  down--going hemisphere. 
This rough estimates indicate that the 
contribution to the muon flux from unflavored mesons  decay 
is in principle observable by 
a neutrino telescope of Km$^3$.

The most interesting scientific goal of a large 
neutrino telescopes such as IceCube is
the detection of astrophysical neutrinos. 
The largest signal from astrophysical neutrinos
could be the isotropic flux from the ensemble of all extragalactic sources. 
The signatures of such an astrophysical neutrino signal are: 
(a) isotropy; (b) a hard energy spectrum; 
(c) approximately equal fluxes of $\nu_e$, $\nu_\mu$ and $\nu_\tau$.
The neutrino fluxes generated by charm decay 
have also the properties (a) and (b), and
equal fluxes for $\nu_e$ and $\nu_\mu$, and therefore constitute a
dangerous background. Most theoretical models for the production 
of astrophysical neutrinos predict an energy spectrum 
harder than what is expected
for the charm decay component, however these predictions 
have important uncertainties

A possible method to separate the astrophysical neutrino 
signal from the charm
decay component is to use measurements of the muon flux to 
constraint the atmospheric neutrino flux, because charm decay 
generates approximately equal fluxes of muons, $\nu_\mu$ and $\nu_e$.
The existence of a dominant muon component from unflavored 
meson decay at very high energy would change critically 
this type of analysis.

\section*{Acknowledgments}
The work of JII and MM has been partially supported by
MICINN of Spain (FPA2010-16802, FPA2006-05294) 
and by Junta de Andaluc\'{\i}a
(FQM 101 and FQM 437).
D.M. was supported by the Deutsche Forschungsgemeinschaft, contract WI
2639/2-1.

\vspace{1.5 cm}

\appendix
\section{Particle Decays}
\label{sec:decays}

In this appendix we describe the lepton spectra produced  
in particle decays that are  used in this  work. The  energy spectrum 
of particle $b$ in the decay of parent particle $a$ can be described in the parent rest 
frame by the function $G_{ab}(x)$, where $x=2E_b^*/m_a$ and $m_a$ is the parent particle mass. 
The spectrum is non vanishing for 
$x$  in the interval [$x_{\rm min}$,$x_{\rm max}$]. 
The lower  limit is $x_{\rm min} = 2 \epsilon$   with $\epsilon=m_b/m_a$,
the upper limit is in general $x_{\rm max} \le 2$ and is  determined by the masses
of the particles in the final state.
The normalization of the functions $G_{ab}$ is chosen to 
be the average multiplicity of particle $b$ in the final state,
\begin{equation}
\langle N_b \rangle = \int_{2\epsilon}^{x_{\rm max}} dx~G_{ab}(x)\, .
\end{equation}
In a frame where $a$ is ultrarelativistic the inclusive 
spectrum of particle $b$ 
takes the scaling form:
\begin{equation}
\frac{dN_{ab}} {dE_b} (E_a, E_b) = \frac{1}{E_a} ~F_{ab} 
\left(y\right)\, ,
\end{equation}
with $y=E_b/E_a$ and
\begin{equation}
F_{ab}(y) = \int_{y + \epsilon^2/y}^2 ~dx~\frac{G_{ab}(x)}
{\sqrt{x^2-4\epsilon^2}}\, .
\label{eq:dec0}
\end{equation}
Equation (\ref{eq:dec0}) assumes that the angular distribution of particle 
$b$ is isotropic in the parent rest frame. This is true for the decay of spin~0 or
unpolarized particles. The decay $Z$--factor is defined as:
\begin{equation}
Z_{ab}(\alpha) = \int_0^1 dy~y^{\alpha-1}~F_{ab}(y)\,.
\end{equation}
In general, particle $b$ can be present in different decay modes. For example $\eta$ 
decays yield muons via the decay modes $\eta \to \mu^+\mu^-\gamma$  and $\eta \to \mu^+\mu^-$. 
The functions $G_{ab}(x)$, $F_{ab}(y)$ and $Z_{ab}(\alpha)$ include a sum over all 
decay channels $j$ where particle $b$ is  produced, weighted by the appropriate
 branching ratios $B_j$. That is:
\begin{equation}
F_{ab}(y) = \sum_{j \in \{ {\rm modes} \} } B_j~F_{ab}^{j}(y)\,,\quad
Z_{ab}(\alpha) = \sum_{j \in \{ {\rm modes} \} } B_j~Z_{ab}^{j}(\alpha)\,,
\end{equation}
with the spectra for each decay mode normalized to unity.

Two--body decays are very simple to treat. In the rest frame the particle spectrum is
 monochromatic, and in an ultrarelativistic frame it is flat between appropriate 
kinematical limits. For the two--body decay of pions and kaons, 
such as $\pi^+ \to \mu^+ \nu_\mu$, one has \cite{gaisser,lipari93}:
\begin{equation}
F_{\pi\mu}(y) = \frac{1}{1-\epsilon^2}~\theta(y-\epsilon^2)\,,\quad
F_{\pi\nu}(y) = \frac{1}{1-\epsilon^2}~\left [1 -\theta(y-1+\epsilon^2)\right]\,,
\end{equation}
with $\epsilon = m_\mu/m_\pi$, and therefore:
\begin{equation}
Z_{\pi\mu}(\alpha) = \frac{1-\epsilon^{2\alpha}}{\alpha(1-\epsilon^2)}\,,\quad
Z_{\pi\nu}(\alpha) = \frac{(1-\epsilon^2)^{\alpha-1}}{\alpha}\,.
\end{equation}

For the decays into a pair of particles with the same mass, 
such as $\eta \to \mu^+ \mu^-$, one has:
\begin{equation}
F_{\eta\mu}^{\mu\mu}(y) = \frac{1}{\sqrt{1-4 \epsilon^2}}
~\left\{\theta[y-y_{\rm min}(\epsilon)]-\theta[y-y_{\rm max}(\epsilon)]\right\}\,,
\end{equation}
with $\epsilon = m_\mu/m_\eta$ and
\begin{equation}
y_{{\rm min},\,{\rm max}}(\epsilon) = \frac{1}{2}
\left(1\mp\sqrt{1-4\epsilon^2}\right)\,.
\label{eq:yminmax}
\end{equation}
The   corresponding $Z$--factor is:
\begin{equation}
Z_{\eta\mu}^{\mu\mu}(\alpha) = \frac{1}{\sqrt{1-4 \epsilon^2}}~\frac{1}{\alpha}
~\left[y_{\rm max}^\alpha-y_{\rm min}^\alpha\right]\,.
\end{equation}

\begin{figure} [ht]
\begin{center}
\includegraphics[width=16.0cm]{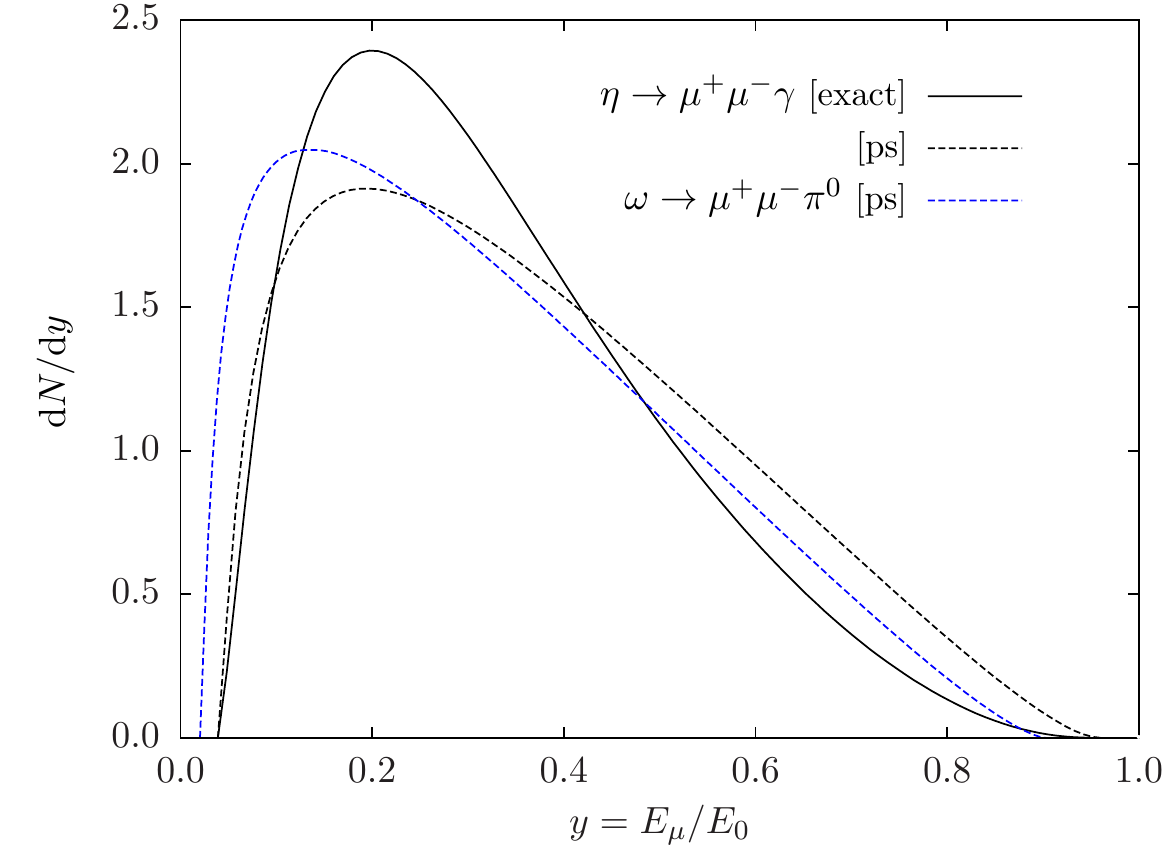}
\end{center}
\caption {\footnotesize
Energy spectra (normalized to unit area) of the muons produced
in  the 3--body decay  of unflavored  mesons.
The solid lines   are for the decay $\eta \to \mu^+ \mu^- \gamma$ 
with (thick) and without (thin) matrix element; the dasheed line
is for the decay $\omega \to \mu^+ \mu^- \pi^\circ$  
(using simple phase space).
\label{fig:decay_spectra} }
\end{figure}

For the decay into three (or more) bodies, the spectrum of the final state particles 
is determined not only by the particle masses, but also by the matrix
element of the decay. As a first approximation one has that the decay of a parent particle 
into three massless final state particles, just from phase space, 
has a spectrum (normalized to unity): 
$G(x) = 2x$, $F(y) =  2(1-y)$ and $Z(\alpha) = 2/(\alpha+ \alpha^2)$.
To estimate the  contribution of unflavored mesons to the muon flux one has to consider 3--body decays such as $\eta \to  \mu^+ \mu^-\gamma$ and $\omega \to  \mu^+ \mu^-\pi^\circ$. The spectrum for the latter has been calculated using phase space. For the decay  $\eta\to \mu^+\mu^-\gamma$, the spectrum without matrix element is:
\begin{equation}
G^{\mu\mu\gamma}_{\omega\mu}(x) = \frac{1}{C_0(\epsilon)}\frac{\left(1-x\right)\sqrt{x^2-4\epsilon^2}}
{1+\epsilon^2-x}\,,
\end{equation}
with $x_{\rm min}=2\epsilon$ and $x_{\rm max}=1$ and the normalization factor
\begin{equation}
C_0(\epsilon) = \frac{1}{2}\sqrt{1-4\epsilon^2}(1+2 \epsilon^2)-\epsilon^2(1-\epsilon^2)\ln\left[\frac{\epsilon^2(1+\sqrt{1-4\epsilon^2})}{1-\sqrt{1-4\epsilon^2}+\epsilon^2(\sqrt{1-4\epsilon^2}-3)}\right ]\,.
\end{equation}
In the ultrarelativistic frame the spectrum becomes
\begin{equation}
F^{\mu\mu\gamma}_{\omega\mu}(y) = \frac{1}{C(\epsilon)}\left\{1-\frac{\epsilon^2}{y}-y+\epsilon^2\ln\left[\frac{\epsilon^2y}{(1-y)(y-\epsilon^2)}\right ]\right\}\,,
\end{equation}
where the kinematical limits $y_{\rm min}$ and $y_{\rm max}$ are again given  by expression (\ref{eq:yminmax}). If the matrix element \cite{Kampf:2005tz} is included then
\begin{equation}
G^{\mu\mu\gamma}_{\omega\mu}(x) = A(x)\frac{\sqrt{x^2-4\epsilon^2}}{(1-x+\epsilon^2)^2}+B(x)
\ln\left[\frac{2\epsilon^2+(1-x)(x+\sqrt{x^2-4\epsilon^2})}
              {2\epsilon^2+(1-x)(x-\sqrt{x^2-4\epsilon^2})}\right]\,,
\end{equation}
with
\begin{eqnarray}
A(x)&=& (1-x)\left[4-11x+7x^2+2(7-8x)\epsilon^2+8\epsilon^4\right]\,,
\\
B(x)&=& 2\left[1-4\epsilon^2-2x(1-x)\right]\,.
\end{eqnarray}
The normalization $C(\epsilon)$ and the distribution in 
the ultrarelativistic frame $F^{\mu\mu\gamma}_{\omega\mu}$ must be evaluated 
numerically (see figure \ref{fig:decay_spectra}).

\begin{figure} [ht]
\begin{center}
\includegraphics[width=16.0cm]{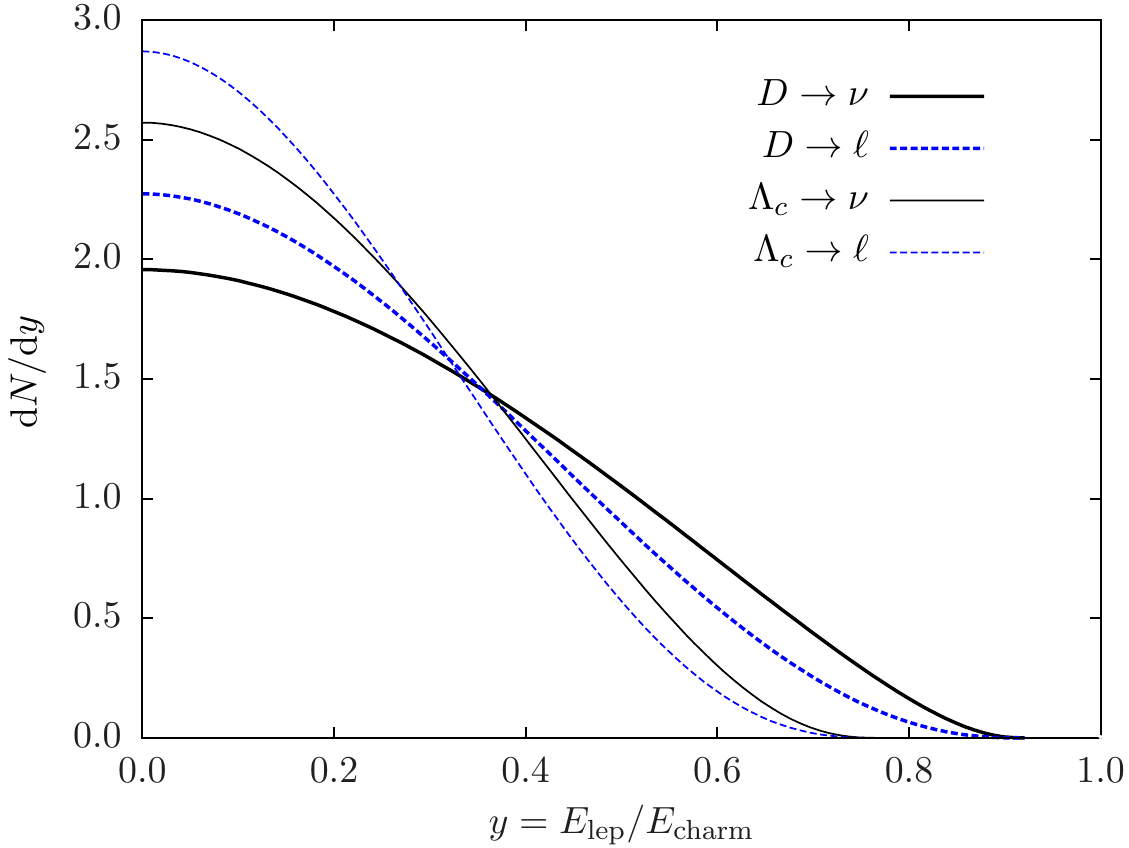}
\end{center}
\caption {\footnotesize
Energy spectra (normalized to unit area) of the muons and neutrinos 
produced in charmed hadron decays.
\label{fig:charmdecay_spectra} }
\end{figure}

\vspace{0.2 cm}
To  describe the decays  of charmed  particles
we have made the simple but reasonably good approximation
to treat the  dynamics of the  charm decay
with the matrix  element at the
the quark level for the decay ($c \to s \ell^+ \nu_\ell)$
but using  the kinematical substitutions:
$M(c) \to M(D)$ and  $M(s) \to M(K)$ for $D$ decay,
and
$M(c) \to M(\Lambda_c)$ and  $M(s) \to M(\Lambda)$ for $\Lambda_c$ decay.
This allows to compute the decay spectra analytically.
Neglecting the  muon mass, in the rest frame of the charmed particle
the muon and neutrino  spectra
can  be written as:
\begin{equation}
G_{ce(\mu)} (x)  =  \frac{1}{C(\epsilon)}
\frac{12x^2\left(1-x-\epsilon^2\right)^2}{1-x}
\end{equation}
and
\begin{equation}
G_{c\nu} (x) = \frac{1}{C(\epsilon)}
\frac{2x^2\left(1-x-\epsilon^2\right)^
     2\left[ 3 - \left( 5 - 2 x \right) x +
      \left( 3 - x \right){\epsilon }^2 \right] }
    {{\left( 1 - x \right) }^3}\,,
\end{equation}
where  $x = 2 E_{\mu,\nu}^*/M(c)$, $\epsilon = M(s)/M(c)$ and
\begin{equation}
C(\epsilon) =
1 - 8{\epsilon }^2 + 8\epsilon^6 -
  {\epsilon }^8 - 12{\epsilon }^4\ln\epsilon^2\,.
\end{equation}
The spectra in a frame where the charmed  particle is
relativistic  can be calculated using (\ref{eq:dec0}):
\begin{equation}
F_{ce(\mu)}(y)
 = \frac{2}{C(\epsilon)}
\left \{ \left( 1 - y - {\epsilon }^2 \right)
     \left[ \left( 1 - y \right)
        \left( 1 + 2y \right)  -
       \left( 5 + 4y \right) {\epsilon }^2 -
       2\,{\epsilon }^4 \right ] +
    6\,{\epsilon }^4\ln \left (\frac{1 - y}{{\epsilon }^2} \right )
    \right \} 
\end{equation}
and
\begin{eqnarray}
F_{c\nu}(y)\!\!&=&\!\! \frac{1}{3C(\epsilon)}\left \{
5-y^2(9-4y)-(27-9y^2)\epsilon^2+\frac{27-9y}{1-y}\epsilon^4-\frac{5-y(4-5y)}{(1-y)^2}\epsilon^6
\right.\nonumber\\&&\left.
+6\epsilon^4(3-\epsilon^2)\ln\frac{1-y}{\epsilon^2}
\right\}\,.
\end{eqnarray}
An important remark is that
the  energy spectra of the charged  leptons
is similar, but slightly softer  than the  corresponding neutrino
spectra (fig.~\ref{fig:charmdecay_spectra}).  From this one can robustly conclude   that  the  expected
fluxes of $\mu^\pm$    from charm decay
  will be   approximately   15--20\% smaller
(depending on the shape of the spectrum and the composition of the
parent charm particles) than the
corresponding $\nu_\mu$ and $\overline{\nu}_\mu$  fluxes. 
The shape of the neutrino and charged  lepton spectra  in 
$D$ and $\Lambda_c$  decay are shown  in fig.~\ref{fig:charmdecay_spectra}.

The calculation of the spectrum of the $\overline{\nu}_\tau$ produced
in  the chain decay  $D_s^+ \to \nu_\tau +\tau^+ \to \nu_\tau+  \overline{\nu}_\tau + X$
should include the effects of the polarization of the 
$\tau^+$.  For the leptonic  modes (such as $\tau^+ \to e^+ \nu_e \overline{\nu}_\tau$
the problem is identical to the case of the chain decay  $\pi \to \mu \to \nu_\mu$.
The details for  the non  leptonic  modes will be  discussed  elsewhere.

\clearpage

\end{document}